# Non-Abelian operation of molecular topological superconductor by n-MOSFET

*Kyoung Hwan Choi[1], and Dong Hack Suh[1†]*

*1 Advanced Materials & Chemical Engineering Building 311, 222 Wangsimni-ro, Seongdong-Gu, Seoul, Korea, E-mail: dhsuh@hanyang.ac.kr*

**Abstract**

Braiding operations are challenging to create topological quantum computers. It is unclear whether braiding operations can be executed with any materials. Although various calculations based on Majorana fermions show braiding possibilities, a braiding operation with a Majorana fermion has not yet been experimentally proven. Herein, braiding operations are demonstrated using a molecular topological superconductor (MTSC) that utilizes the topological properties intrinsic in molecules. The braiding operations were implemented by controlling two MTSC modules made by pelletizing crystals of 4,5,9,10-tetrakis(dodecyloxy)pyrene, which is proposed as the first MTSC material through n-MOSFETs. It shows the elements of topological quantum computers that can be demonstrated without an external magnetic field at room temperature.

# Introduction

Topological quantum computation (TQC) is an approach to fault tolerant quantum computation, in which unitary quantum gates result from the braiding of certain topological quantum objects called 'anyons'.[1,2] Anyons can arise in two ways: as localized excitations of an interacting quantum Hamiltonians[3] or as defects in a system.[4] Not all anyons are useful in topological quantum computation; only non-Abelian anyons are applicable, and these do not include anyonic excitations that are believed to occur in most odd-denominator fractional quantum Hall states. A collection of non-Abelian anyons at fixed positions and with fixed local quantum numbers has a non-trivial topological degeneracy. This topological degeneracy allows quantum computation as braiding enables unitary operations between the distinct degenerate states of the system. The unitary transformations resulting from braiding depend only on the topological classification of the braid, thereby endowing them with fault tolerance.

Exotic excitations obeying non-Abelian statistics play a crucial role in TQC and can emerge in various condensed matter systems[5,6], including chiral p-wave superfluids, and superconductors[7,8]. Such non-Abelian quasiparticles are believed to be realized in the A phase of superfluid $^3$He[9], the oxide superconductor $Sr_2RuO_4$[10], p-wave superfluids in cold atom settings[11], as well as at the interfaces of s-wave superconductors and topological insulators[12]. Certain vortex excitations in these systems support zero-energy Majorana fermions residing in their cores, which leads to the topological degeneracy of ground states and non-Abelian statistics when spatially separated multiple vortices are present. The final state is determined by the topology of the braiding and is robust against local perturbation, thereby having possible applications for topological quantum computation.[13]

Recently, the experimental realization of topological superconductors originated from polyaromatic hydrocarbon-based molecules.[14] Their conductance showed a constant conductance plateau and a zero-bias conductance peak, which simultaneously depicts the Majorana hinge and non-local Majorana zero modes(MZMs) of odd parity. A strategy to realize the non-Abelian braiding-like operation on chiral Majorana fermions coupled with a quantum dot (QD) or MZM was suggested.[15]

Here, the experimental results of braiding operations of molecular topological superconductors (MTSCs) coupled with an n-MOSFET are reported. This additional experimental evidence proves that aromatic molecules contain Majorana fermions and are the best strategy for realizing topological quantum computers at room temperature without an external magnetic field.

## Results and discussion

In previous studies, experiments on the Hall conductivity of crystals of the orthorhombic phase of 4,5,9,10-tetrakis (dodecyloxy) pyrene (TDP) proved the non-trivial characteristics of polyaromatic hydrocarbons (PAHs) as intrinsic topological superconductors.[14,16] In particular, the coherent quantum phase slip (CQPS), the constant conductivity, and the zero-bias conductance peak confirmed that non-trivial properties intrinsic in PAHs can be realized in mesoscopic physics without extremely low temperatures and external magnetic fields. It shows new possibilities in the study of topological superconductors. A topological superconductor material with such properties has been proposed as a molecular topological superconductor (MTSC). Here, braiding operations by linking two MTSC modules would be experimentally demonstrated.

TDP, which was proposed as a molecule capable of implementing the first MTSC, was synthesized according to the existing method and then characterized through NMR and IR (Figure. S1 - S3). It was crystallized according to the previously reported method, and the

crystal structure was characterized by XRD. (Figure. S4)

In the case of an MTSC, there is an intrinsic spin current due to the magnon embedded in the molecule, so it was essential to properly control the intrinsic current of the PAH. (Figure. 1) The most important idea here was that both electrons and holes should be allowed to move in order to maintain the Majorana fermion. In a study of splitting of a Cooper pair, it was possible to stably produce results by forming a coherent phase by using the quantum dot, and it was recently reported that a current in chiral quantum Hall insulators can also make a braiding-like operation using the quantum dot. However, since an MTSC already had a complete Majorana fermion pair inside the molecule, it was always able to stably implement a coherent phase. In other words, even without the coherent phase and tuning using the quantum dot, self-coherent characteristics of an MTSC can be transmitted through general conductors. This is the reason why braiding operations can be performed in an MTSC with an n-MOSFET instead using the quantum dot. Furthermore, when $V_{SG}$ becomes larger than $V_{th}$, the trivial state is expressed, and the non-trivial state is diminished.

In this manuscript, the braiding operation of an MTSC was proved in two configurations. The first configuration can verify whether the n-MOSFET performs the on/off function of the non-trivial state between two MTSC modules. When $V_{SG} < V_{th}$, the MTSC module, MTSC modules 1 and 2 are not connected through an n-MOSFET, but are composed of the edge state of Majorana fermions, located at the ends of the wire. (Figure. 2a) In this case, it can be expected that $\gamma_0$ and $\gamma_1$ are entangled with each other to show the characteristics of the Majorana fermions depending on $V_{SD}$. On the other hand, if the n-MOSFET acquires a chemical potential larger than $V_{th}$ and a current is generated in the loop due to electron transfer, the edge mode disappears. (Figure. 2b) However, if $E_{f,i}$ is located in the surface band gap of an MTSC during the $V_{SG}$ sweep, a non-trivial state implemented only by electrons can exist throughout the loop.

(Figure. 2e) This is because the n-MOSFET only allows the flow of electrons.

Furthermore, braiding operations between two MTSC modules were demonstrated by controlling the connectivity of the edge mode with the n-MOSFET. The second configuration was formed by connecting one more lead to the MTSC in the module 1. In this case, if $V_{SG} < V_{th}$, $\gamma_0$ and $\gamma_1$ or $\gamma_0$ and $\gamma_2$ become entangled. (Figure. 2c) However, as $V_{SG}$ gradually increases and the n-MOSFET turns on, the trivial state is activated between $\gamma_0$ and $\gamma_1$, and it is expected that $\gamma_2$ becomes an isolated Majorana fermion. (Figure. 2d)

The experiment was conducted through Keithly 4200SCS, and $V_{th}$ of the n-MOSFET was 2.0V. (Figure. S5) The current was measured through the configuration corresponding to Figure. 2a and 2b. $V_{SD}$ was set from -0.55mV to -0.05mV in 0.05mV steps, and $V_{SG}$ swept from 2V to 3.4V in 5mV steps. The first thing to be verified was that the current gradually increases and becomes saturated as $V_{SG}$ increases. (Figure. 3a) In addition, the relationship between the saturated current and $V_{SD}$ is almost perfectly linear when $V_{SD}$ is from -0.55mV to -0.2mV. This result suggests two important things. First, the saturated current obeys Ohm's Law. (Figure. S6) Therefore, the saturated current can be estimated as having the trivial characteristics of electrons.

The second important point is that the n-MOSFET performs the on/off functions for the non-trivial and trivial states accurately. The initial current change that starts to open when $V_{SG} > V_{th}$ is applied is clearly verified. When $V_{SD}$ less than -0.45mV, the current increases linearly in the initial specific section and then exponentially as $V_{SG}$ increases. Interestingly, this linearly increasing current is confirmed in a broader $V_{SG}$ section as the $V_{SD}$ decreases. When $V_{SD}$ are -0.30, -0.25, and -0.2mV, the linear current region can be clearly identified. If the x-intercept is obtained by extrapolating the region of the exponentially increasing current, it can be considered that the linear current increase is not due to trivial electron characteristics. (Figure.

S7-S9)

The current measured at $V_{SD}$ = - 0.15mV shows a sawtooth profile, which is the signature of the gap inversion at a topological transition and the unique oscillatory pattern that originates from Majorana interference.[17,18] The sawtooth profile has been predicted in researches related to Majorana fermions, but has not yet been demonstrated experimentally. According to the previous study, it was experimentally confirmed that the intrinsic PAH current performs NNN hopping only at a specific chemical potential,[19] and there was also a statement that NNN hopping implies a non-trivial state of PAHs.[20] Therefore, the sawtooth profile can be thought of as resulting from the intrinsic non-trivial properties of the molecule.

The reason this phenomenon occurs only at specific $V_{SD}$ is more clearly demonstrated through the differential conductivity. (Figure. 3b) The most remarkable thing about the differential conductivity is that a conductance peak cannot be confirmed at -0.55mV, whereas the negative conductance peak is proved between -0.5mV and -0.25mV. These peaks can be understood as the alignment between the Majorana zero modes of MTSCs and $E_{f,i}$ of the n-MOSFET. Therefore, only electrons among the Majorana fermions inherent in the MTSC module 1 are transferred to the MTSC module 2 on the fermi surface of the n-MOSFET. Although the MTSC is a DIII material, the n-MOSFET can only allow electrons to move. So, it exhibits the same dip pattern in the differential conductance as a topological insulator. When $V_{SD}$ is -0.25mV, a slightly more interesting phenomenon is noticed. The pattern of the peaks identified in the difference conductance has positive as well as negative peaks because the non-local conductance has an odd function conductivity near the gap of the closing-reopening topological phase transition point.[21]

At $V_{SD}$ = -0.15mV, a positive conductance peak is only identified. This phenomenon was interpreted as occurring because the chemical potential of the circuit by $V_{SD}$ entirely entered the surface band gap of the MTSC and $E_{f,i}$ was completely matched by $E_F$. If $V_{SD}$ has a value outside the surface band gap, electrons are likely to be transferred, even if a potential barrier is created. Moreover, if $V_{SD}$ has a value inside the surface band gap, electrons will be transferred if any potential barrier exists. Therefore, both electrons and holes can flow only at $V_{SD}=-0.15mV$ with $V_{SG} = 3.145V$. This is reminiscent of the dip in the peak conductance transition, in which the topological insulator becomes a topological superconductor.

Through transport spectroscopy in which the n-MOSFET is arranged between the MTSC modules, it was demonstrated that the amount of current could be controlled by $V_{SD}$ and that it was possible to switch between the trivial and non-trivial states through the control of $V_{SG}$. Due to this characteristic, braiding operations can be demonstrated through the changes of the currents $I_1$ and $I_2$ according to $V_{SG}$ and $V_{SD}$ in the second configuration. Comparing the currents $I_1$ and $I_2$ measured at $V_{SD} = -0.6mV$, it is verified that $I_1$ gradually decreases and $I_2$ gradually increases as $V_{SG}$ becomes larger than $V_{th}$. (Figure. 4a)

In the first configuration of this experiment, the edge state of the Majorana fermion was verified according to $V_{SG}$ and, if $V_{SG} < V_{th}$, the edge state appeared; if $V_{SG} > V_{th}$, a trivial state was developed by the flow of electrons. Likewise, $I_2$ showed a linear current increase at the low $V_{SG}$ and an exponential increase from $V_{SG} = 2.018V$. This means that $\gamma_0$ and $\gamma_1$ disappear. Accordingly, $I_1$ shows an oscillating current as an isolated Majorana fermion. These characteristics are maintained to $V_{SD} = -0.25mV$. (Figure. S10 - S12) These characteristics can be proved more clearly through the differential conductance. The differential conductance at $I_2$ has a negative conductance peak at high $V_{SD}$ (Figure. 4b), and it can be proved that a positive conductance peak is increased as $V_{SD}$ is decreased. (Figure. S13 - S15)

At $V_{SD} = -0.20mV$, a completely new current pattern that can convey a deep meaning to the operation of the Majorana fermion was identified. In this case, as $V_{SD}$ exceeds $V_{th}$, $I_1$ decreased, and $I_2$ increased. (Figure. 4c) Furthermore, a sawtooth pattern similar to the first configuration was verified. The sawtooth pattern in the second configuration demonstrates that the currents generated in the first and second loops are complementary. As the current increases in $I_2$, the current decreases in $I_1$. More interestingly, the current in $I_1$ maintains an oscillating pattern. This shows that some $\gamma_2$ are entangled with $\gamma_1$, but most $\gamma_2$ remain isolated.

The results of the second configuration can also be understood more clearly through the differential conductance. When $V_{SD}$ is -0.2mV, the differential conductance of $I_1$ has an almost constantly oscillating differential conductance regardless of $V_{SG}$. On the other hand, the differential conductance of $I_2$ has a very small positive conductance peak at $V_{SG}$ = -2.08V, but has a very strong positive differential conductance peak as the $V_{SG}$ becomes larger.

A differential conductance peak of $I_1$ shows odd-parity, while a differential conductance peak of $I_2$ shows even-parity. The shape of peaks seen in the differential conductance gives important implications for the isolation and connection of Majorana fermions. In general, isolated Majorana fermions have odd-parity differential conductance peaks, confirming the entanglement of $\gamma_0$ and $\gamma_1$ and the isolation of $\gamma_2$.

**Conclusion**

It was confirmed that braiding operations between two modules of MTSCs based on PAHs, can be controlled. In the experiment in which two MTSC modules were connected through the n-MOSFET, it was demonstrated that the chemical potential of the MTSC can control the transfer of non-trivial states through the control of $V_{SD}$. In addition, the transition between non-trivial and trivial states through the control of $V_{SG}$ was verified. By connecting one lead to the MTSC module 1, it was confirmed that a braiding operation is possible between the MTSC modules 1 and 2 depending on $V_{SG}$. This is the first result showing the possibility of forming a quantum computer at room temperature without an external magnetic field.

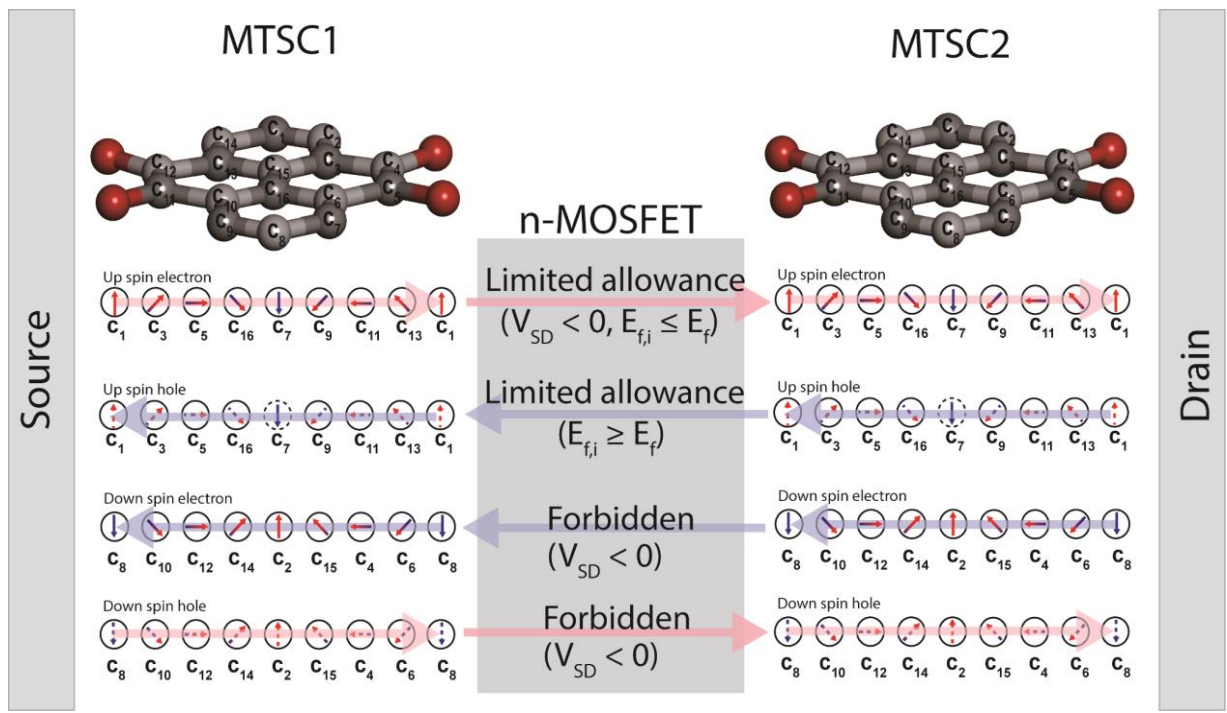

**Figure 1. Braiding operation of molecular topological superconductors (MTSCs) with n-MOSFET.** It was confirmed that MTSCs have intrinsic spin currents that perform NNN hopping in different directions within the molecule based on up spin and down spin. When connecting two MTSCs with n-MOSFET, the way that electrons and holes of up spin and down spin are transferred will be different depending on $V_{SD}$ and $V_{SG}$. For example, if $V_{SD}$ is negative, only up spin electrons and holes can be used as charge carriers, and electrons and holes selectively become charge carriers according to $V_{SG}$. This principle makes it possible to implement braiding operation with MTSCs.

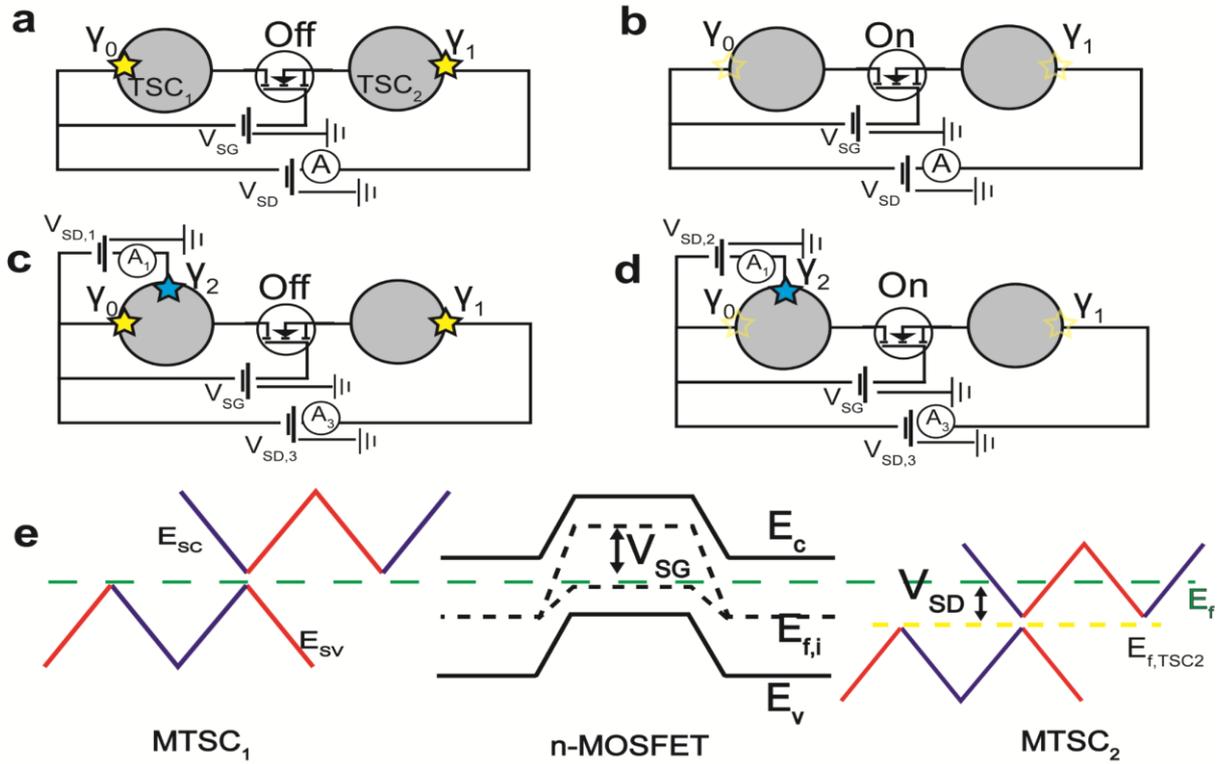

**Figure 2. Schematic diagram of MTSC- based braiding operation experiment and band structure. a and b.** A circuit connecting n-MOSFET and one lead to each of the two MTSCs to form one loop. When $V_{SG}<V_{th}$ (a), edge states of Majorana fermions of $\gamma_0$ and $\gamma_1$ can be formed at the ends of MTSC modules 1 and 2. On the other hand, if $V_{SG}>V_{th}$(b), the edge state disappears because all circuits are connected. **c and d.** A circuit connecting n-MOSFET and two leads of MTSC module 1, and one lead of MTSC module 2 to form two loops. The second configuration can experiment with braiding operation. Loop 1 is always connected regardless of $V_{SG}$. On the other hand, loop 2 can adjust the edge state according to the $V_{SG}$. **e.** band structure of MTSC and n-MOSFET

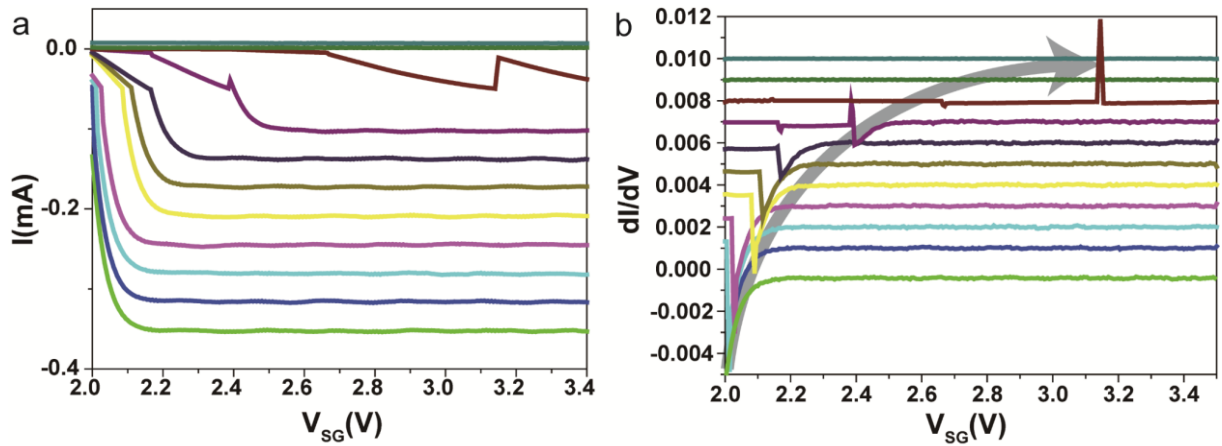

**Figure 3. Transport spectroscopy for the first configuration at various $V_{SD}$. a.** IV curves of first configuration. **b.** Differential conductances of first configuration.

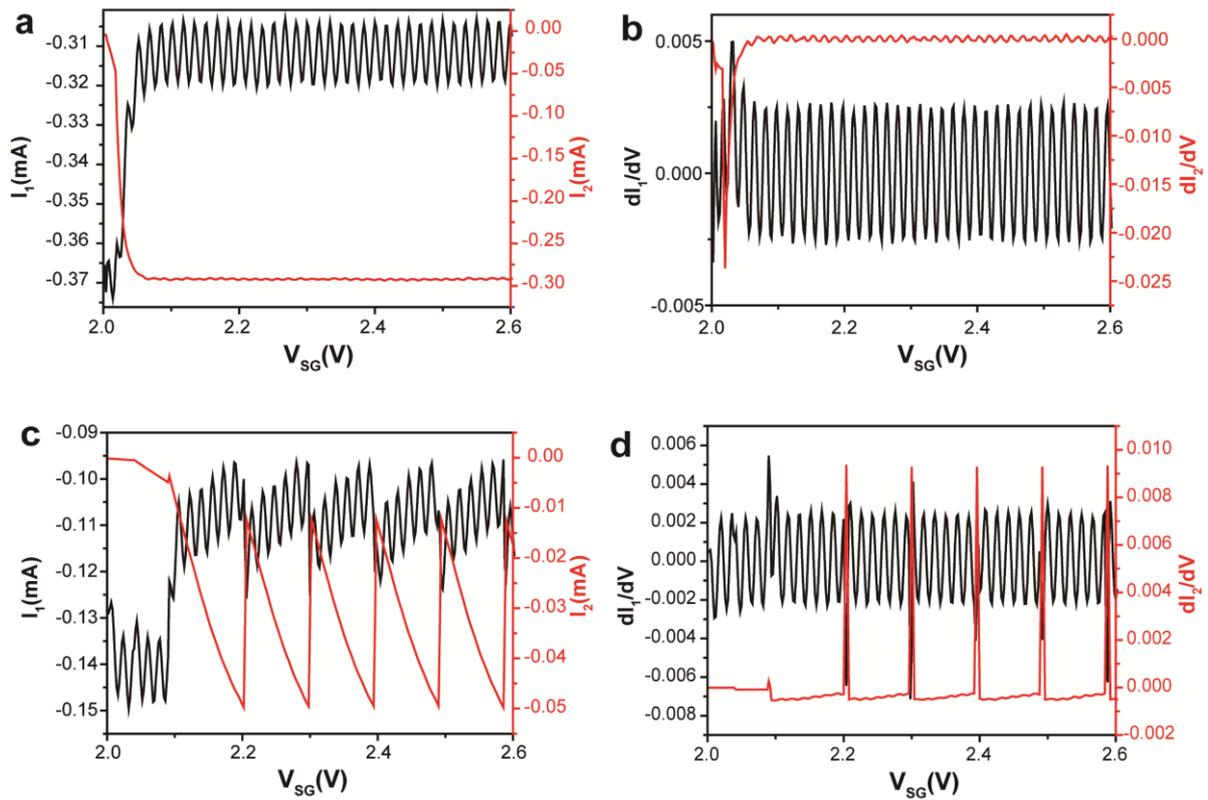

**Figure 4. Transport spectroscopy for the second configuration. a and b.** IV curves and differential conductances of second configuration at $V_{SD}$ = -0.4mV. **c and d.** IV curves and differential conductances of second configuration at $V_{SD}$ = -0.2mV.

# Non-Abelian operation of molecular topological superconductor by n-MOSFET

*Kyoung Hwan Choi[1], and Dong Hack Suh[1†]*

*1 Advanced Materials & Chemical Engineering Building 311, 222 Wangsimni-ro, Seongdong-Gu, Seoul, Korea, E-mail: dhsuh@hanyang.ac.kr*

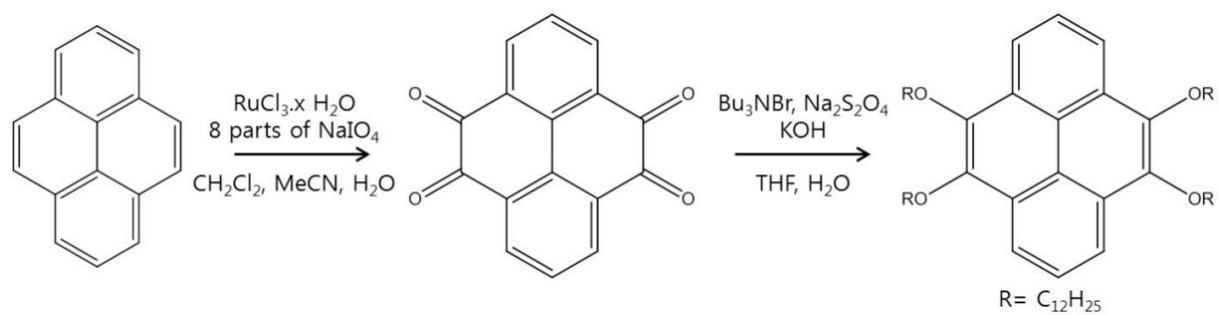

**Figure S1.** Synthesis of 4,5,9,10-tetrakis(dodecyloxy)-pyrene.

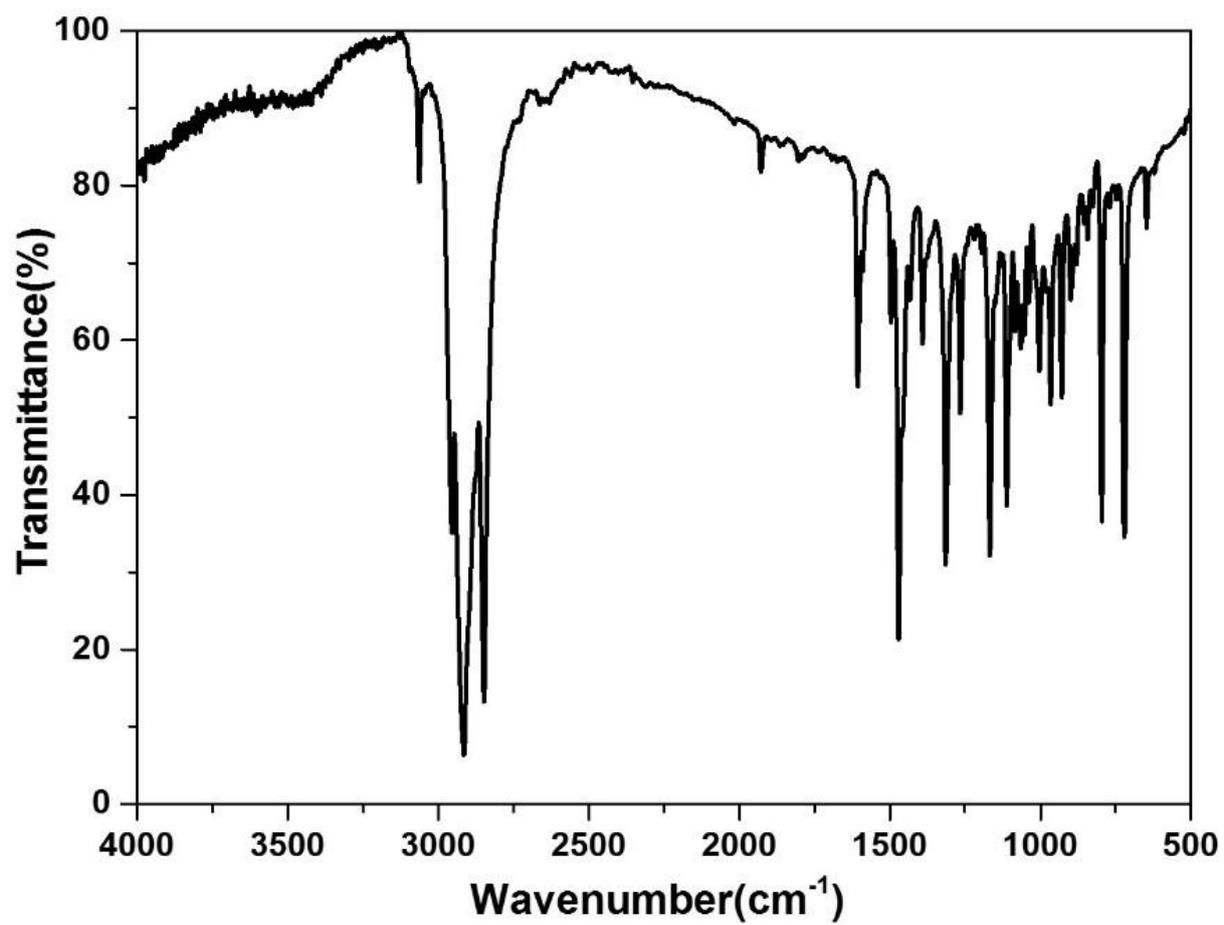

**Figure S2.** Infrared spectra of 4,5,9,10-tetrakis(dodecyloxy)-pyrene.

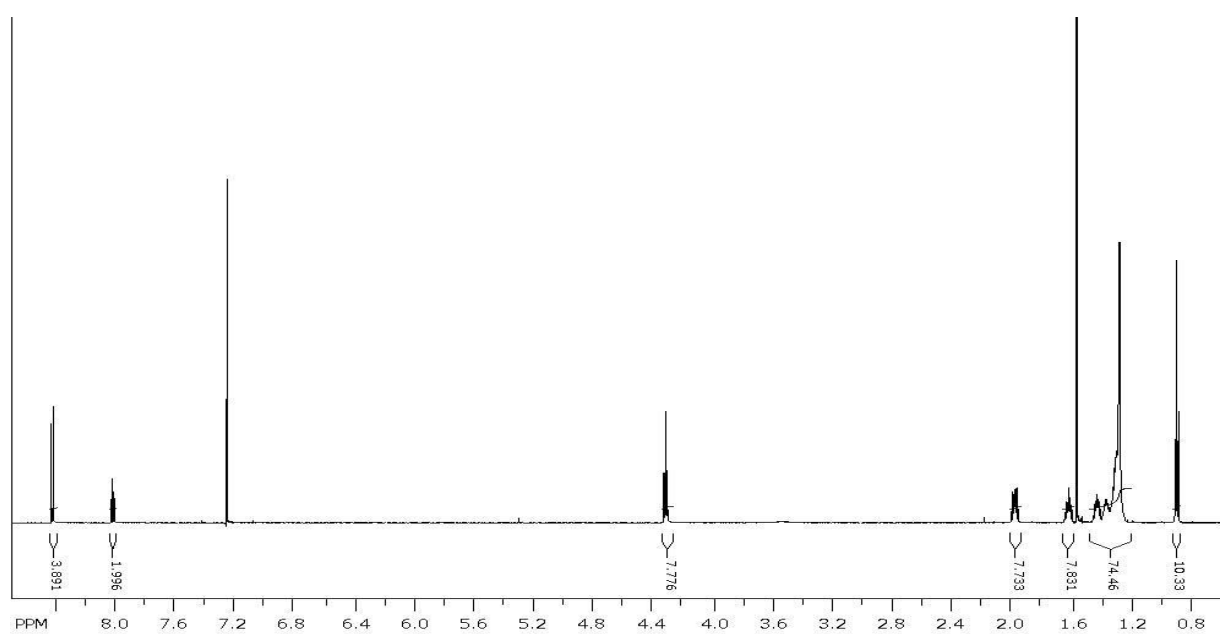

**Figure. S3** The NMR characterization of 4,5,9,10-tetrakis(dodecyloxy)-pyrene. 1H NMR (600MHz, CDCl3) δ 8.32(d, 4H), 7.71(t, 2H), 4,21(t, 8H), 1.91(m, 8H), 1.57(m, 8H), 1.40-1.27(m, 80H), 0.88(t, 12H).

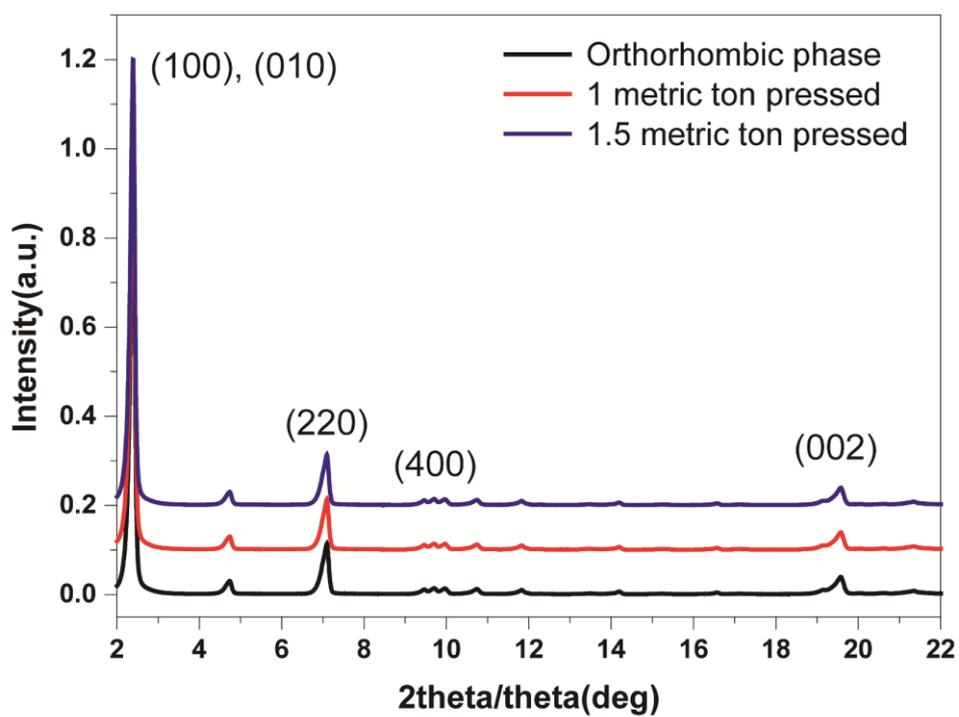

**Figure. S4.** XRD patterns of pelletized orthorhombic phase of TDP.

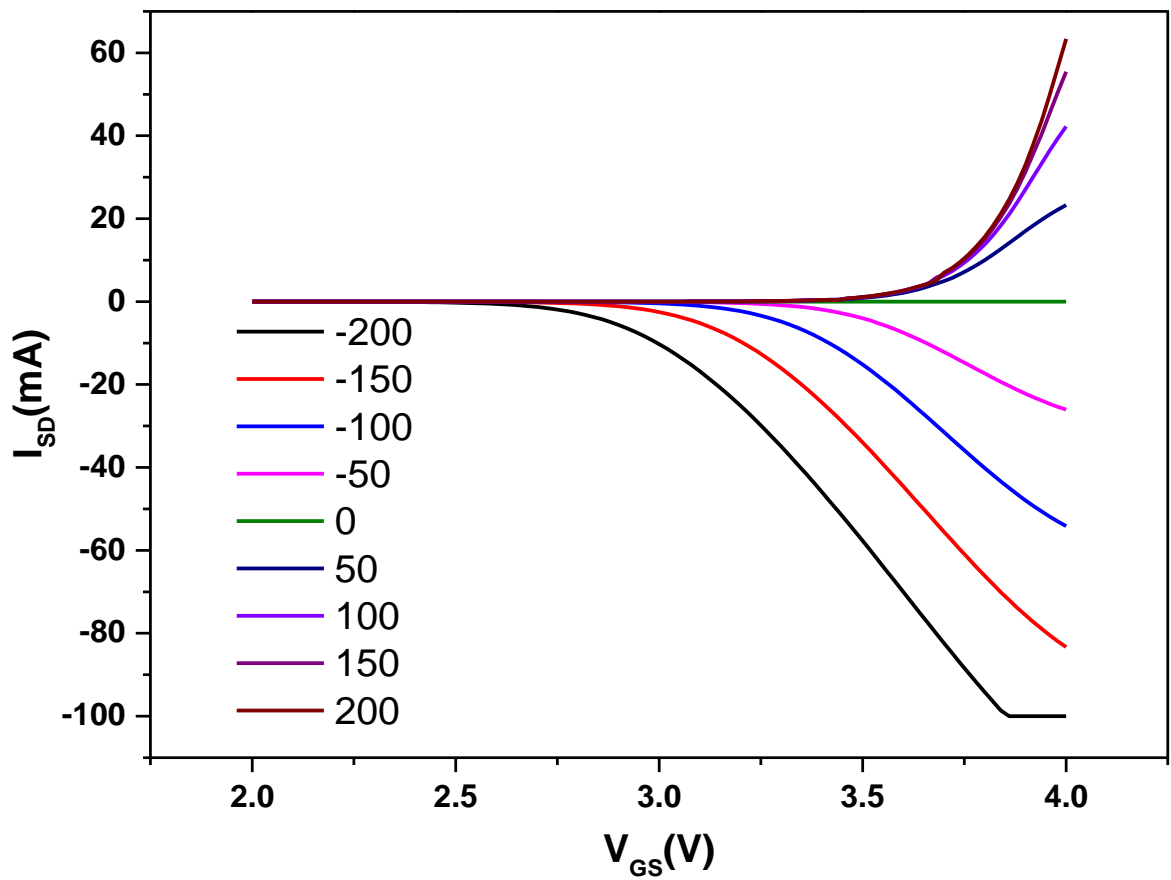

Figure S5. IV curve of n-MOSFET at various $V_{SD}$.

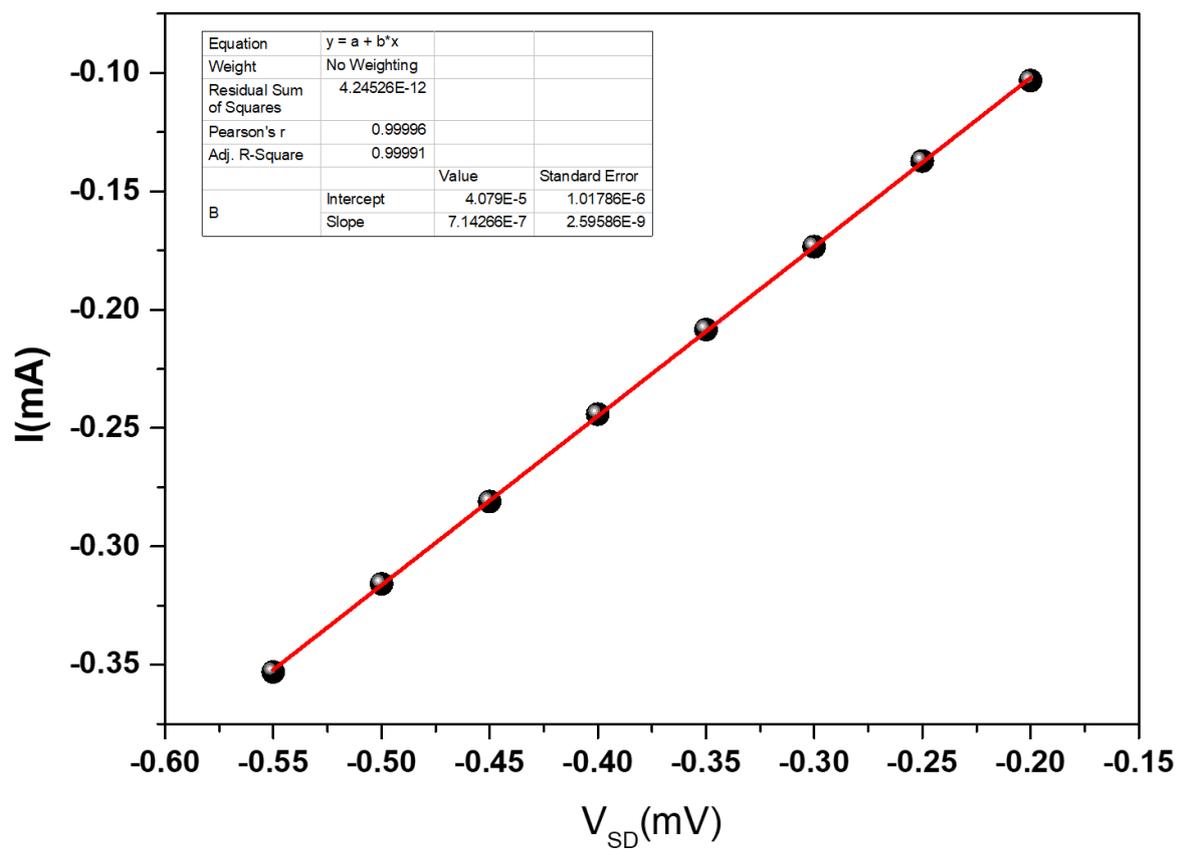

**Figure S6.** Saturated current of between MTSC modules at various $V_{SD}$.

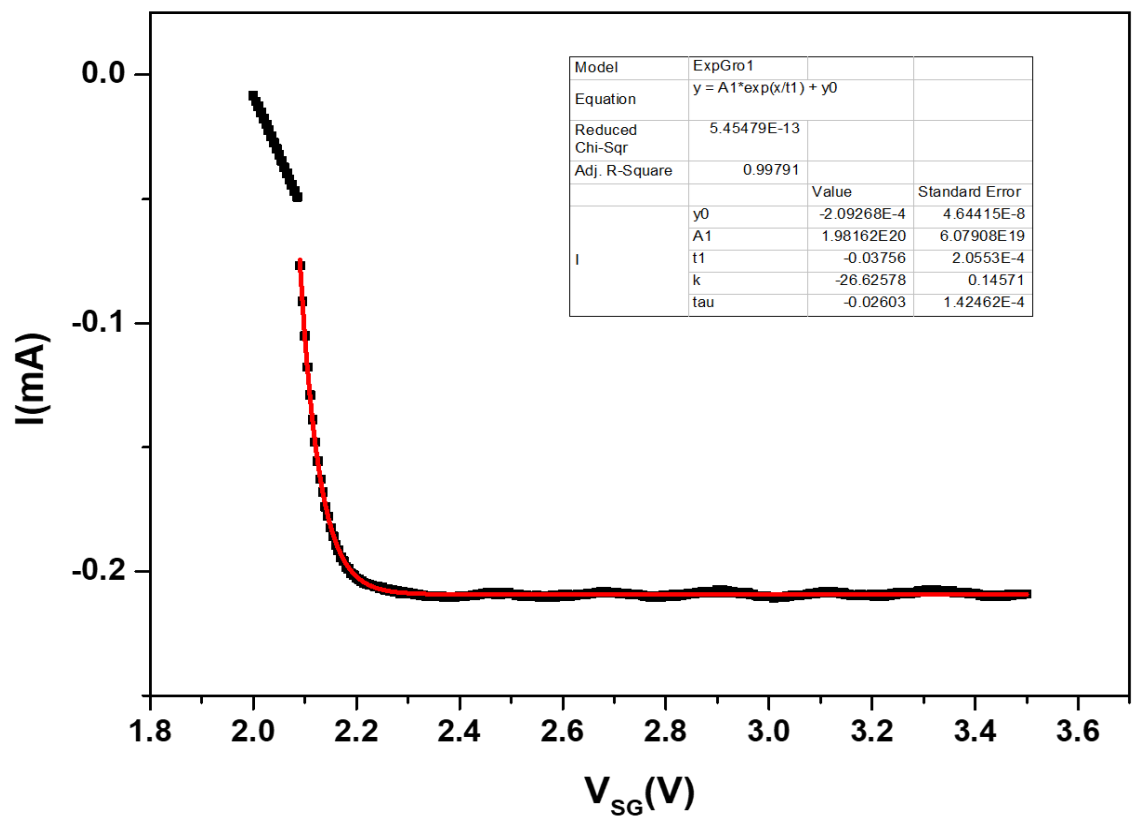

Figure S7. Exponential curve fitting of current of MTSC modules at $V_{SD}$ = -0.35mV. According to the fitting results, the x-intercept is $V_{SG}$ = 2.07 V

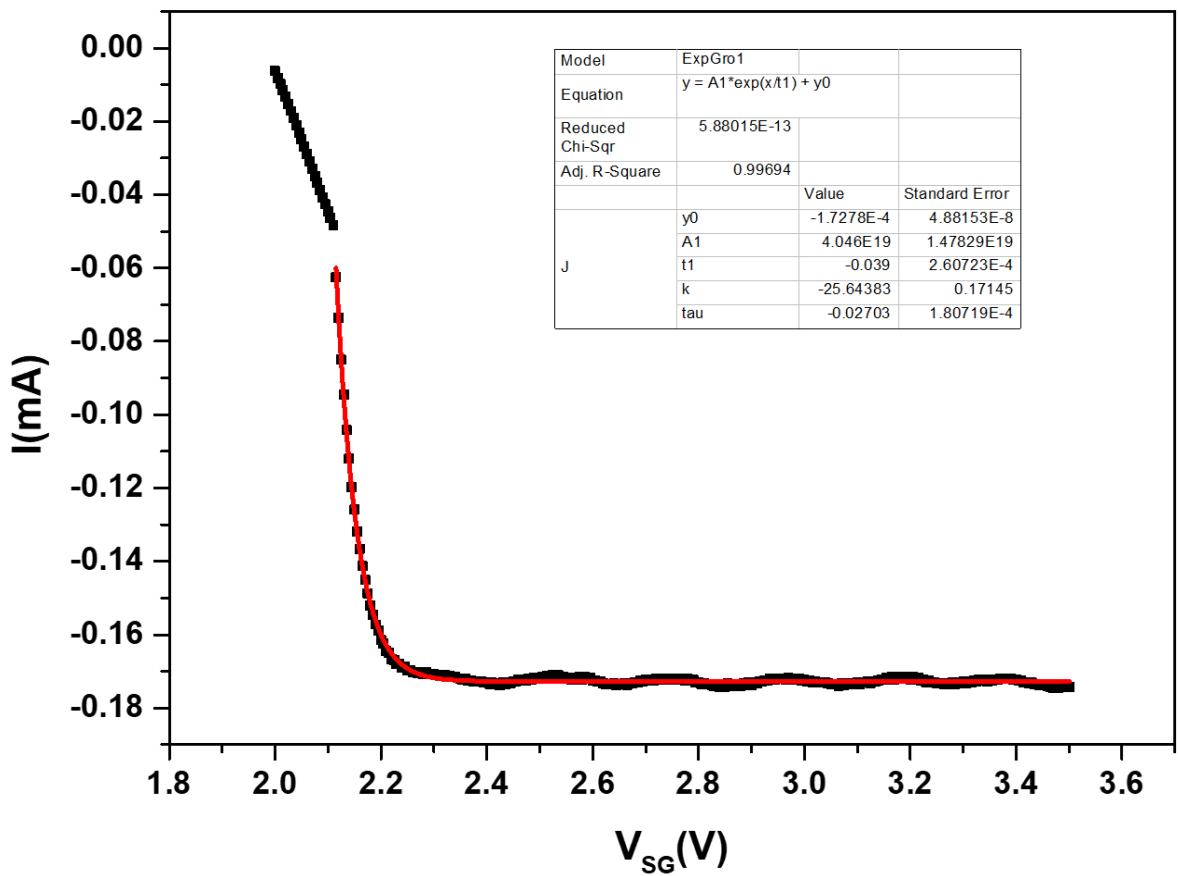

Figure S8. Exponential curve fitting of current of MTSC modules at $V_{SD}$ = -0.30mV. According to the fitting results, the x-intercept is $V_{SG}$ = 2.10 V

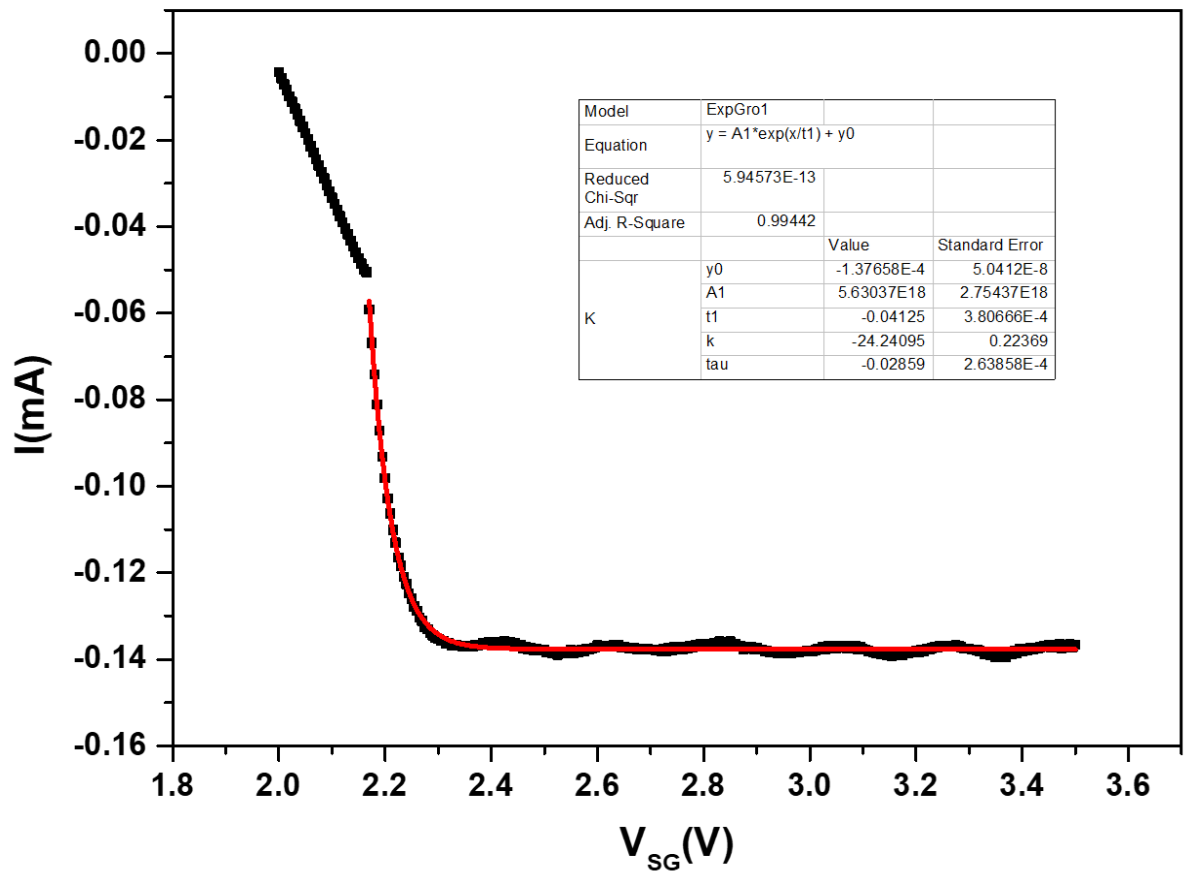

Figure S9. Exponential curve fitting of current of MTSC modules at $V_{SD}$ = -0.25mV. According to the fitting results, the x-intercept is $V_{SG}$ = 2.15 V

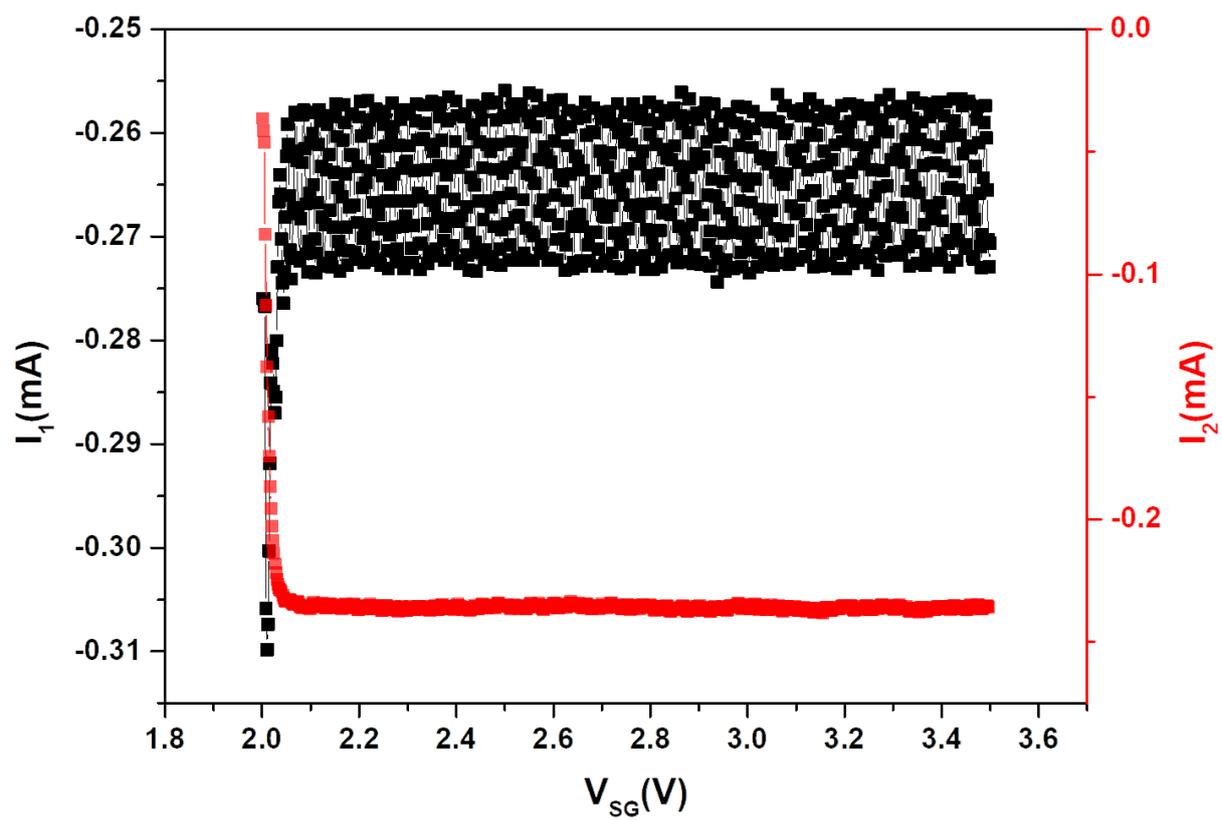

**Figure S10.** IV curve of second configuration at $V_{SD}$ = -0.5mV

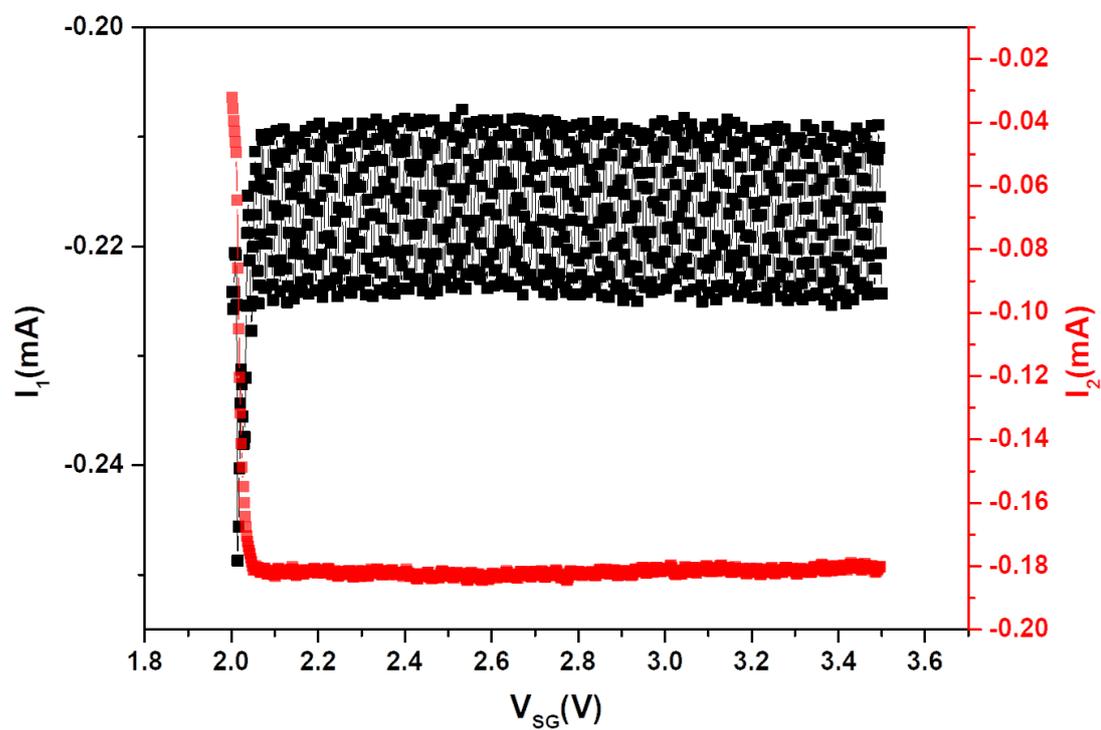

**Figure S11.** IV curve of second configuration at $V_{SD}$ = -0.4mV

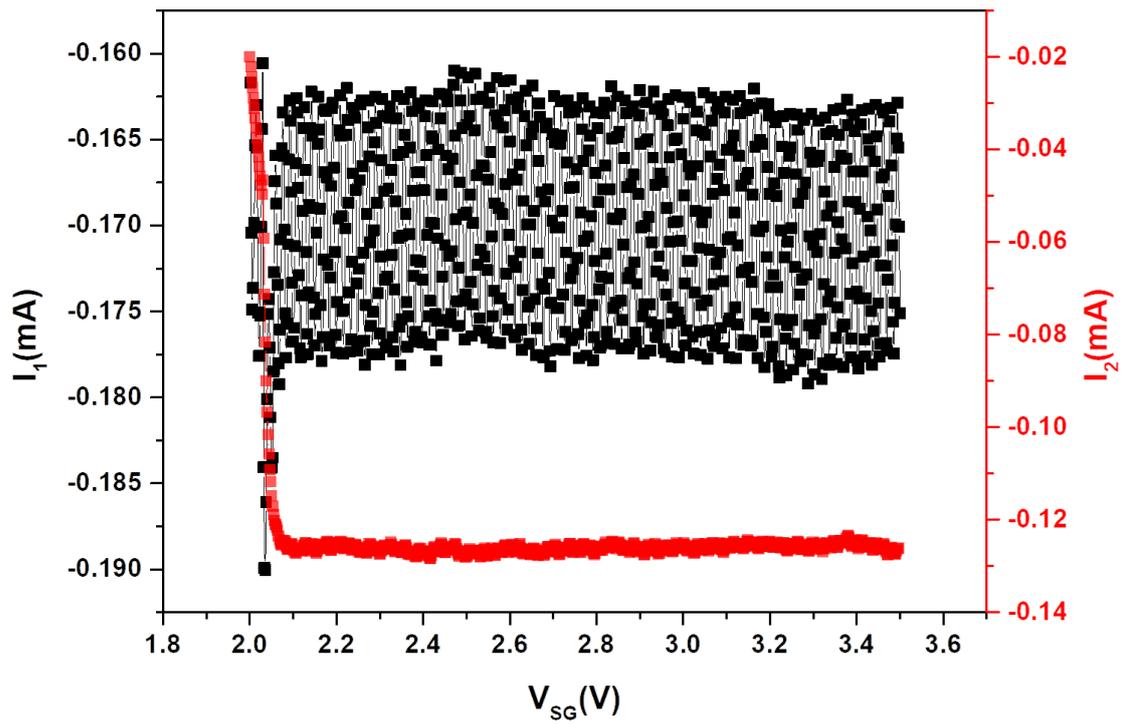

**Figure S12.** IV curve of second configuration at $V_{SD}$ = -0.3mV

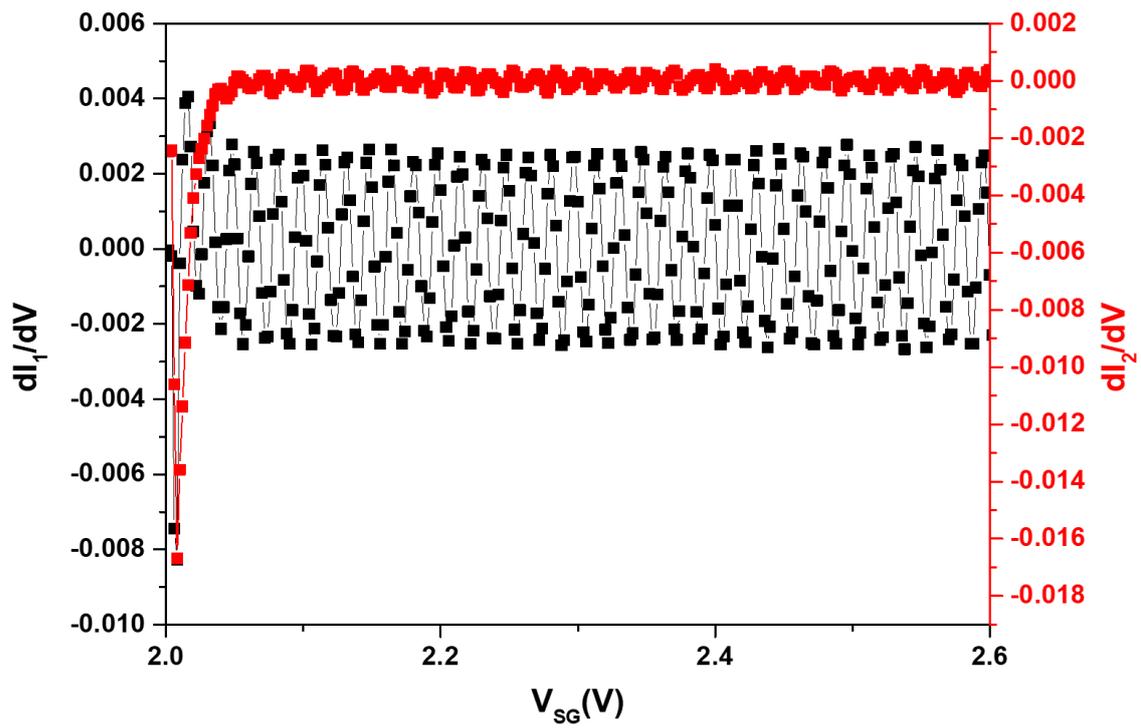

**Figure S13.** Differential conductance curve of second configuration at $V_{SD}$ = -0.5mV

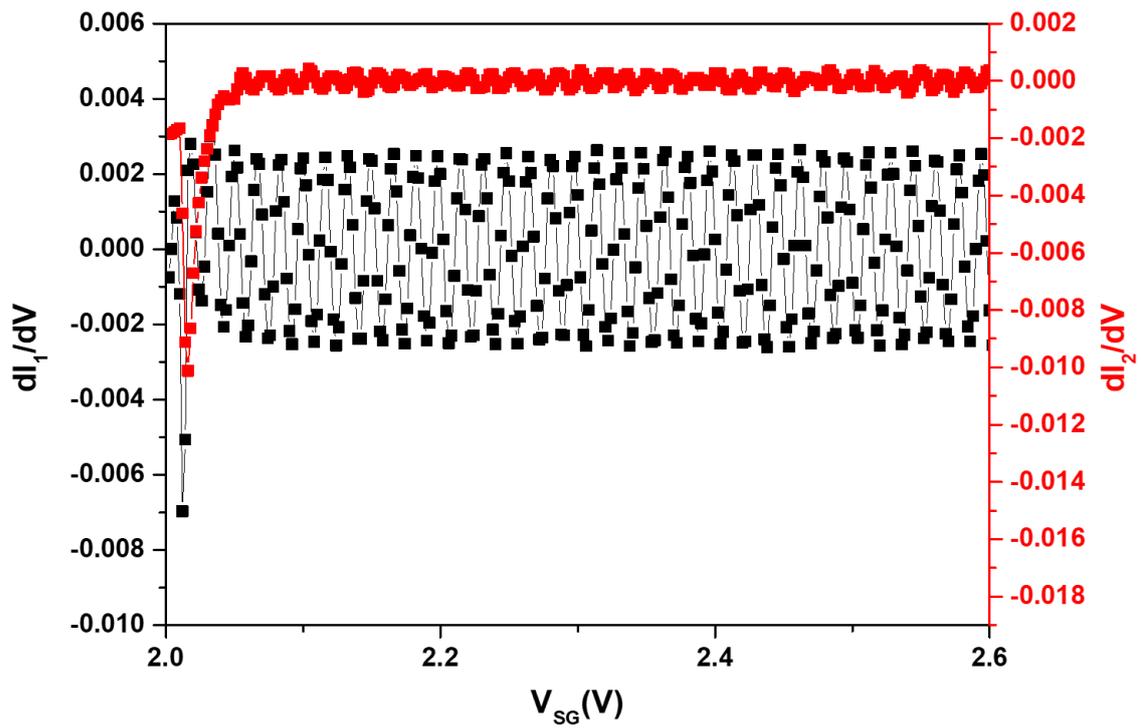

**Figure S14.** Differential conductance curve of second configuration at $V_{SD}$ = -0.4mV

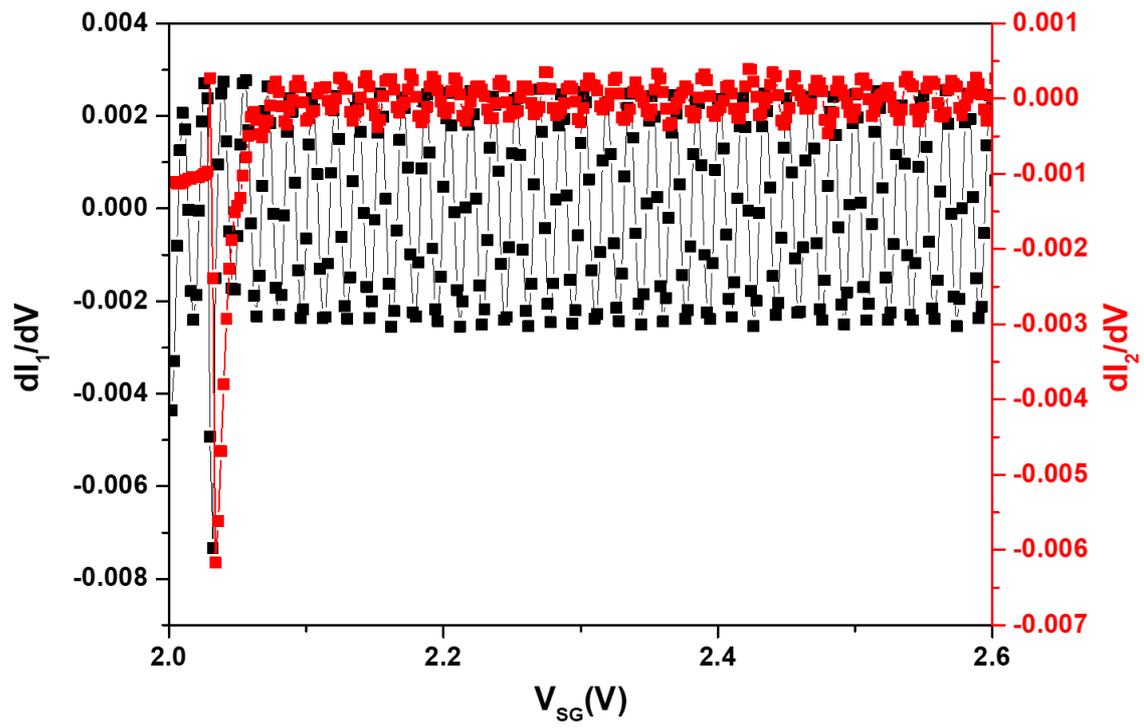

**Figure S15.** Differential conductance curve of second configuration at $V_{SD}$ = -0.3mV